\documentclass[prl,twocolumn,superscriptaddress,floatfix,noshowpacs,10pt,longbibliography]{revtex4-1}%
\usepackage{graphicx,bm,times}
\usepackage{amsmath}
\usepackage{amsfonts}
\usepackage{amssymb}

\usepackage{bm}
\usepackage{color}
\usepackage{times}

\usepackage{xcolor}
\usepackage[%
  colorlinks=true,
  urlcolor=blue,
  linkcolor=blue,
  citecolor=blue
  ]{hyperref}

\begin{document}

\title{Anomalous high-magnetic field  electronic state of the nematic superconductors FeSe$_{1-x}$S$_x$}

\author{M. Bristow}
\affiliation{Clarendon Laboratory, Department of Physics,
University of Oxford, Parks Road, Oxford OX1 3PU, UK}

\author{P. Reiss}
\affiliation{Clarendon Laboratory, Department of Physics,
University of Oxford, Parks Road, Oxford OX1 3PU, UK}

\author{A. A. Haghighirad}
\affiliation{Clarendon Laboratory, Department of Physics,
University of Oxford, Parks Road, Oxford OX1 3PU, UK}
\affiliation{Institut fur Festk\"{o}rperphysik, Karlsruhe Institute of Technology, 76021 Karlsruhe, Germany}

\author{Z. Zajicek}
\affiliation{Clarendon Laboratory, Department of Physics,
University of Oxford, Parks Road, Oxford OX1 3PU, UK}

\author{S. J. Singh}
\affiliation{Clarendon Laboratory, Department of Physics,
University of Oxford, Parks Road, Oxford OX1 3PU, UK}

\author{T. Wolf}
\affiliation{Institut fur Festk\"{o}rperphysik, Karlsruhe Institute of Technology, 76021 Karlsruhe, Germany}

\author{D. Graf}
\affiliation{National High Magnetic Field Laboratory and Department of Physics, Florida State University, Tallahassee, Florida 32306, USA}

\author{W. Knafo}
\affiliation{Laboratoire National des Champs Magn\'etiques Intenses (LNCMI-EMFL),
UPR 3228, CNRS-UJF-UPS-INSA, 143 Avenue de Rangueil, 31400 Toulouse, France}

\author{A.\;McCollam}
\affiliation{High Field Magnet Laboratory (HFML-EMFL), Radboud University, 6525 ED Nijmegen, The Netherlands}

\author{A. I. Coldea}
\email[corresponding author:]{amalia.coldea@physics.ox.ac.uk}
\affiliation{Clarendon Laboratory, Department of Physics,
University of Oxford, Parks Road, Oxford OX1 3PU, UK}

\begin{abstract}
Understanding superconductivity requires detailed knowledge of the normal electronic state from which it emerges.
A nematic electronic state that breaks the rotational symmetry of the lattice can potentially promote
unique scattering relevant for superconductivity. Here,  we investigate the normal transport of superconducting FeSe$_{1-x}$S$_x$  across a nematic
phase transition using high magnetic fields up to 69~T to establish the temperature and field-dependencies.
We find that the nematic state is an anomalous non-Fermi liquid, dominated by a linear resistivity at low temperatures
that can transform into a Fermi liquid, depending on the  composition $x$ and the impurity level.
Near the nematic end point, we find an extended temperature regime with  $\sim T^{1.5}$ resistivity.
The transverse magnetoresistance inside the nematic phase has as  a $\sim H^{1.55}$ dependence over a large magnetic field range
and it displays an unusual peak at low temperatures inside the nematic phase.
Our study reveals anomalous transport inside the nematic phase,
driven by the subtle interplay between the changes in the electronic structure of
a multi-band system and the unusual scattering processes affected by large magnetic fields and disorder.

\end{abstract}
\date{\today}
\maketitle

Magnetic field is a unique tuning parameter
that can suppress superconductivity
to reveal the normal low-temperature electronic behavior of
many unconventional superconductors \cite{Ramshaw2015,Boebinger1996}.
High-magnetic fields can also induce new phases of matter,
probe Fermi surfaces and determine the quasi-particle masses from quantum oscillations
in the proximity of quantum critical points \cite{Ramshaw2015,Coldea2019}.
In unconventional superconductors, close to antiferromagnetic critical regions,
an unusual scaling between a linear resistivity in temperature and magnetic fields was found
\cite{Hayes2016,Giraldo-Gallo2018}.
Magnetic fields can also induce metal-to-insulator transitions,
as in hole-doped cuprates,
where superconductivity emerges from an exotic electronic ground state \cite{Boebinger1996}.

FeSe is a unique bulk superconductor with $T_c \sim 9$~K which
displays a variety of complex and competing electronic phases \cite{Coldea2017}.
FeSe is a bad metal at room temperature and
it enters a nematic electronic state below $T_s \sim 87$~K.
This nematic phase is characterized by multi-band shifts
driven by orbital ordering that lead to Fermi surface distortions \cite{Coldea2017,Watson2015a}.
Furthermore, the electronic ground state is that of a strongly correlated system
and the quasiparticle masses display orbital-dependent enhancements \cite{Watson2017a,Watson2015a}.
FeSe shows no long-range magnetic order at ambient pressure, but complex magnetic fluctuations are present
 at high energies over a large temperature range \cite{Wang2016b}.
 Below $T_s$, the spin-lattice relaxation rate from NMR experiments is enhanced
 as it captures the low-energy tail of  the stripe spin-fluctuations \cite{Kasahara2016,Shi2018}.
 Furthermore, recent $\mu$SR studies invoke the close proximity of FeSe to a magnetic quantum critical point
 as the muon relaxation rate shows unusual temperature dependence inside the nematic state \cite{Grinenko2018}.

The changes in the electronic structure and magnetic fluctuations of FeSe can have profound
implication on its transport and superconducting properties. STM reveals a highly anisotropic superconducting gap
driven by orbital-selective Cooper pairing \cite{Sprau2016}.
Due to the the presence of the small inner bands, whose Fermi energies are  comparable to the superconducting gap,
FeSe was placed inside the BCS-BEC crossover regime \cite{Kasahara2014}. In large magnetic fields, when the Zeeman energy
is comparable to the gap and Fermi energies,
a peculiar highly-polarized superconducting state may occur \cite{Kasahara2014}.

To establish the role played by different competing interactions on nematicity and superconductivity,
an ideal route is provided by the isoelectronic substitution of selenium by sulphur ions in FeSe$_{1-x}$S$_x$
\cite{Watson2015c}. This tuning parameter suppresses nematicity and it leads to changes in the electronic structure, similar to the temperature effects,
with the Fermi surface becoming isotropic in the tetragonal phase and the electronic correlations
becoming weaker \cite{Watson2015c,Coldea2017,Coldea2019,Reiss2017}.
As nematicity is suppressed, it creates ideal conditions to explore a potential nematic critical point \cite{Hosoi2016}
in the absence of magnetism. The superconducting dome extends outside the nematic state
but anisotropic pairing remains robust  \cite{Sato2018} and a different superconducting state
was suggested to be stabilized in the tetragonal phase \cite{Hanaguri2018}.

\begin{figure*}[htbp]
	\centering
\includegraphics[trim={0cm 0cm 0cm 0cm}, width=1\linewidth,clip=true]{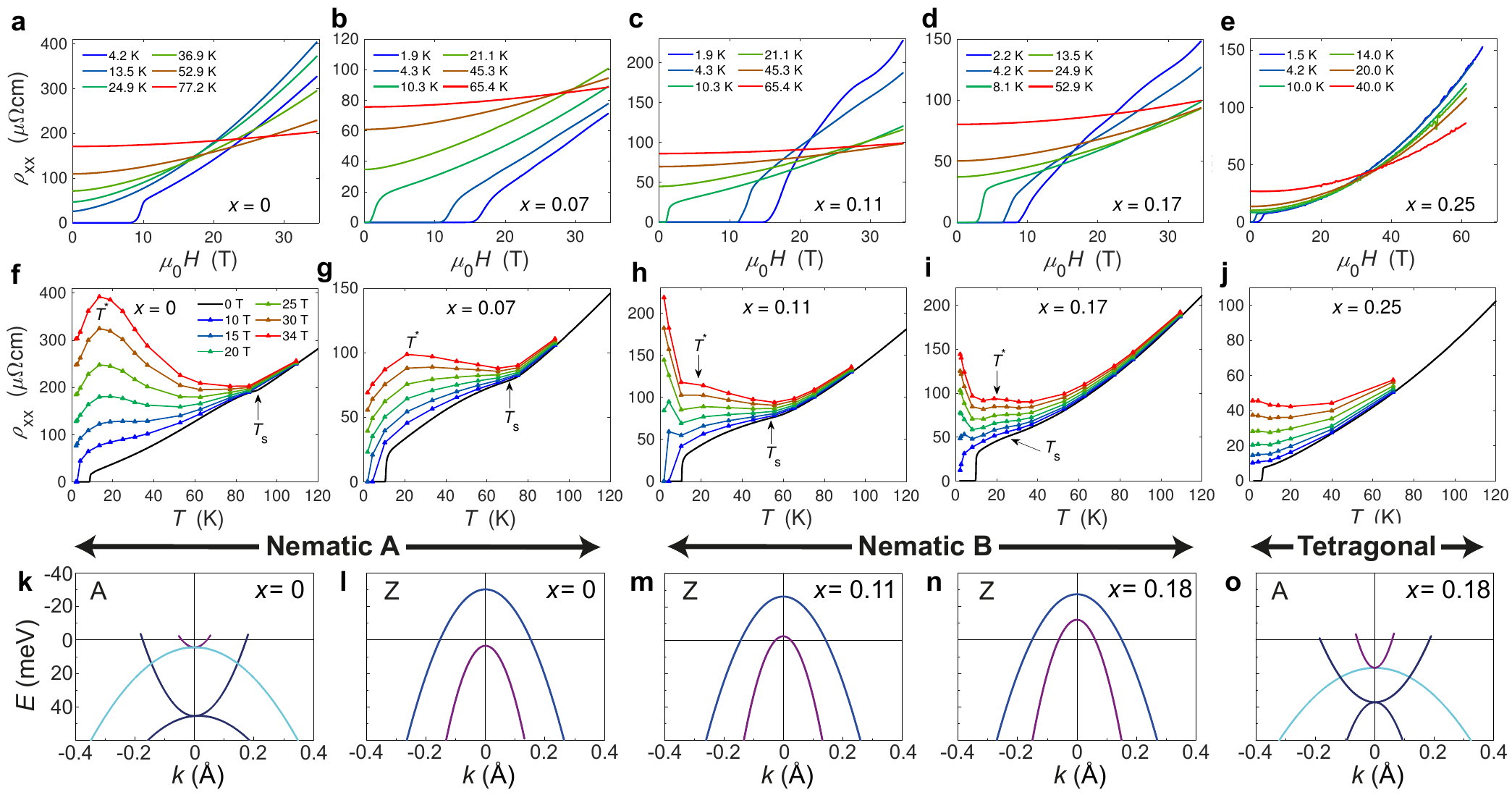}
	\caption{\textbf{Transverse magnetoresistance of the nematic and tetragonal FeSe$_{1-x}$S$_x$.}
(a-e) Field-dependent in-plane resistivity  at different constant temperatures for different compositions, $x$, inside
and outside the nematic phase. The magnetic field is applied along the $c$-axis, perpendicular
to the in-plane electrical current. A strong magnetoresistance develops inside the nematic phase.
(f-j) Resistivity against temperature in zero field (solid line) and at fixed magnetic fields (symbols), as extracted from the top panel
 for different $x$. The peak in magnetoresistance is indicated by $T^*$  and the nematic phase emerges at $T_s$.
 (k-o) Schematic band dispersion at low temperatures at two high symmetry points at the top of the Brillouin zone, $Z$ and $A$
 for different $x$ (based on ARPES data reported in Refs.\onlinecite{Watson2015a,Watson2015c,Reiss2017,Coldea2017}).
 The horizontal lines represent the location of distinct regions in the  magnetotransport behaviour called nematic A ($x=0, 0.07$),
 nematic B ($x=0.11, 0.17$) and the tetragonal phase for $x \gtrsim 0.18$. In the tetragonal phase, the compensated semi-metal is
formed of two electron and two-hole like bands. Deep inside the nematic phase  the inner hole band and inner electron
bands are brought in the vicinity of the Fermi level.}
	\label{fig1}
\end{figure*}

In this paper we study the normal electronic state across the nematic transition
in FeSe$_{1-x}$S$_x$ using magnetotransport studies in high-magnetic fields up to 69~T.
 We find that the nematic state
 has a non-Fermi-liquid behaviour with an unusual transverse magnetoresistance ($\sim H^{1.55}$),
reflecting an unconventional scattering mechanism.
Just outside the nematic phase, resistivity is dominated
by a $\sim T^{1.5}$ dependence, similar to studies under pressure \cite{Reiss2019}.
The transverse magnetoresistance is significant inside the nematic phase
and it shows an unusual change in slope at low temperatures.
Inside the nematic phase at low temperatures, we find linear resistivity
followed by Fermi-liquid behaviour for certain $x$ and impurity levels.
Our study reveals anomalous transport in the nematic state
due to the subtle changes in the electronic structure and/or scattering, which are
also influenced by impurity levels.

\vspace{-0.7cm}
\section{Results and Discussion}
\vspace{-0.5cm}
Figs.~\ref{fig1}a-e show the transverse magnetoresistance, $\rho_{xx}$,
 of different single crystals of  Fe(Se$_{1-x}$S$_x$) up to 35~T  at various fixed temperatures
 inside the nematic phase and up to 69~T for $x\sim 0.25$ in the tetragonal phase.
 From these constant temperature runs,
 we can extract the magnetoresistance at fixed fields for each composition $x$, as shown  in Fig.~\ref{fig1}f-j,
which reveals several striking features.
Firstly,  the magnetoresistance
increases significantly once a system enters the nematic state at $T_s$,
and its magnitude dependents on the concentration $x$,
being largest for FeSe, just above $T_c$.
\begin{figure}[htbp]
	\centering
\includegraphics[trim={0cm 0cm 0cm 0cm}, width=1\linewidth,clip=true]{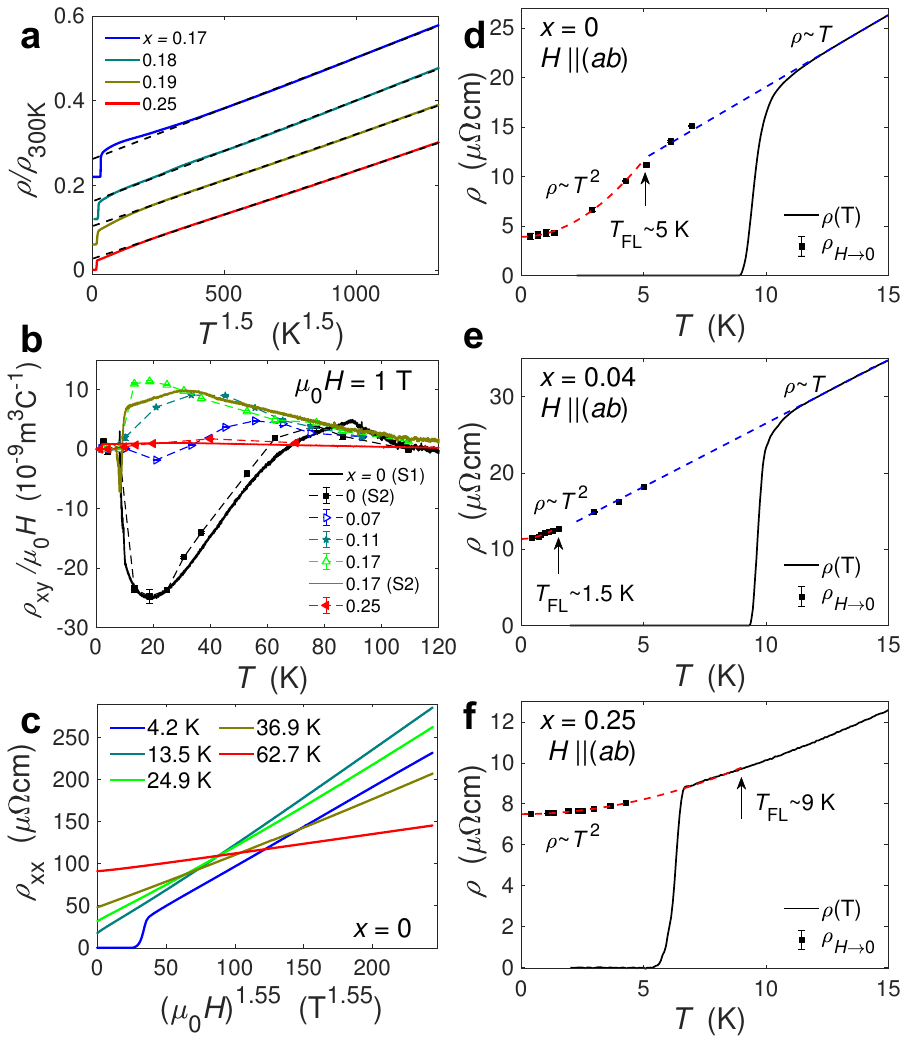}
	\caption{\textbf{Normal electronic state of FeSe$_{1-x}$S$_{x}$.}
(a) Temperature dependence of resistivity versus $T^{1.5}$ over a large temperature
region just outside the nematic phase.
(b) Hall effect coefficient in low magnetic fields, indicating the change in sign
and the dominance of different highly mobile carriers across the nematic phase.
(c) Resistivity versus $H^{1.55}$  for FeSe inside the nematic phase at constant temperatures.
(d-f) The low-temperature linear resistivity. The solid lines are the zero-field resistivity data.
Solid circles represent the zero-field extrapolated values of $\rho_{xx}$ when $H||(ab)$ plane.
The dashed lines represent fits to  a Fermi-liquid behaviour found below $T_{FL}$,
as indicated by arrows.}
	\label{fig2}
\end{figure}
Secondly,  in the vicinity of $T_c$ in magnetic fields much larger than the upper critical field,
the magnetoresistance shows an unusual temperature dependence that varies strongly
with $x$ across the phase diagram, as shown in  Fig.~\ref{fig1}(f-g).
The resistivity slope $d \rho_{xx}/dT$ in  34~T of FeSe changes sign around a crossover temperature, $T^* \sim $ 14~K,
as shown in Fig.\ref{fig1}f  (also in the colour plot of the slope in Fig.~\ref{fig3}d).
With increasing sulphur substitution from FeSe towards $x\sim 0.07$ (defined as the nematic A region),
the position of $T^*$ shifts to a slightly higher temperature  of $\sim 20$~K, and the peak in magnetoresistance
is much smaller than for FeSe.
For higher concentrations, approaching the nematic phase boundary,
($x  \sim 0.11 -0.17$ defined as the nematic B region),
there is a small peak at $T^*$ but the
 negative slope $d \rho_{xx}/dT$  in 34~T is strongly enhanced
at low temperatures,  different from the nematic A phase (see Fig.~\ref{fig1}(h,i) and Fig.~\ref{fig3}(d)).
Lastly, in the tetragonal phase, the magnetoresistance shows a conventional behaviour
and increases quadratically in magnetic fields (Fig.~\ref{fig1}(e) and (j)).

The unusual downturn in resistivity in  high-field fields
below $T^*$  inside the nematic A phase was previously  assigned
 to large superconducting fluctuations in FeSe in magnetic fields up to 16~T \cite{Kasahara2016,Shi2018}.
We find that this behaviour remains robust in magnetic fields
 at least a factor of 2 higher than the upper critical field of $ \sim $16~T for $H||c$ \cite{Kasahara2016}.
Furthermore, it also manifests in $x \sim 0.07$ inside the nematic A phase
but it disappears for higher $x \gtrsim 0.1$.
As $T_c$ and the upper critical field
inside the nematic phase for different $x$
remain close to that of FeSe \cite{Coldea2019,Bristow2019Hc2},
the changes in the resistivity slope in high magnetic fields
are likely driven by field-induced effects that influence scattering and/or the electronic structure.

The Hall coefficient, $R_H = \rho_{xy}/ \mu_0 H$, extrapolated in the low-field limit (below 1~T)
for FeSe$_{1-x}$S$_x$ has an unusual temperature dependence, as shown in Fig.\ref{fig2}b.
For a compensated metal, the sign of the Hall coefficient depends on the difference
between the hole and electron mobilities  \cite{Watson2015b}.
In the tetragonal phase above $T_s$ and for $x\gtrsim 0.18$,  $R_H$ is close to zero (Fig.\ref{fig2}b),
as expected for a two-band compensated metal.
On the other hand,  in the  low-temperature nematic A phase the sign of $R_H$ is negative suggesting
that transport is dominated by a highly mobile electron band \cite{Watson2015c,Sun2016mob}.
It becomes positive inside the nematic B phase,
dominated by a hole-like band (Fig.~\ref{fig2}(a)). It is worth mentioning
that  inside the nematic B phase the quantum oscillations are dominated by
a low-frequency pocket with light-mass that disappears at the nematic end point \cite{Coldea2019}.
Thus, the behaviour of $R_H$ is linked to the disappearance of a
 small 3D hole pocket center at the $Z$-point in FeSe below $T_s$ and its re-emergence
in the nematic B phase with $x$ substitution around $x \sim 0.11$,
as found in ARPES studies \cite{Watson2015c} and sketched in  Fig.\ref{fig1}(m).
Interestingly, the subtle changes in the electronic structure in FeSe$_{1-x}$S$_x$ seem to correlate with the
different features observed both in magnetoresistance  (Fig.\ref{fig1}(f-i))
and in the Hall coefficient $|R_H|$  that shows a maximum near $T^*$ (Fig.\ref{fig2}(b)).
In a high magnetic field, the Hall component of FeSe is complex,
changing sign and being non-linear \cite{Watson2015c,Bristow2019Hc2}.
A magnetic field can induce
changes in scattering and/or field-induced Fermi-surface effects
in the limit when the cyclotron energy is close to the Zeeman energy.
The smallest inner  bands of FeSe$_{1-x}$S$_x$
shift in energy as a function of composition $x$ (and temperature \cite{Coldea2019}), as shown in Figs.\ref{fig1}(k-o).
Furthermore, Hall effect in iron-based superconductors can be affected
by the spin fluctuations that induce mixing of the electron and hole currents
\cite{Fanfarillo2012}.

Next, we attempt to quantify the magnetoresistance
across the phase diagram and in the vicinity of the nematic end point
 in FeSe$_{1-x}$S$_{x}$, as shown in Fig.\ref{fig1}(a-e).
At the lowest temperature, inside the nematic phase,
 the transverse magnetoresistance of most samples is dominated by quantum oscillations \cite{Coldea2019}
making difficult to quantify its dependence. A near-linear magnetoresistance
is detected for $x\sim 0.07$ in Fig.~\ref{fig1}b and
 for a dirty sample (with low residual resistivity ratio $\sim 8.5$) in Fig.~\ref{FigSM_extrapolation_comparison_dirty}.
The quasi-linear field magnetoresistance at low temperature can arise from squeezed
trajectories  of  carriers  in  semiclassically  large  magnetic  fields in case of small Fermi surfaces
($\omega_c \tau  \gg 1$) \cite{Pippard1989,Du2005}.
Another explanation for an almost linear magnetoresistance
is the presence of mobility fluctuations caused by spatial inhomogeneities, as found in
low carrier density systems \cite{Narayanan2015,Du2005,Singleton2018}.

Classical magnetoresistance in systems with a single dominant scattering time
is expected to follow a $H^2$ dependence  \cite{Pippard1989}.
This results in Kohler's rule,
which is violated in FeSe$_{1-x}$S$_x$ suggesting that the magnetoresistance is not dominated
 by a single scattering time,  as shown in Fig.~\ref{FigSM_MR_Scaling}(a-c).
Magnetoresistance
 is quadratic in magnetic fields up to 69~T
 in the tetragonal phase ($x \geq 0.19$) (see Fig.\ref{fig1}e and Fig.\ref{FigSM_Hscaling_1p55_all}(e-f))
 but not inside the nematic phase.
FeSe$_{1-x}$S$_x$ are compensated multi-band systems \cite{Coldea2017}
where the high-field magnetoresistance
is expected to be very large and dependent on
 scattering times of electron and hole bands \cite{Watson2015b}.
 Magnetoresistance has a complex form and instead simpler scaling have been
sought to reveal its importance, in particular in the vicinity of critical points \cite{Hayes2016,Giraldo-Gallo2018}.
 For example, in BaFe$_2$(As$_{1-x}$P$_x$) for $x \sim 0.33$ at the antiferromagnetic critical point,
 a universal $H-T$ scaling was empirically found between the linear resistivity in temperature and magnetic field \cite{Hayes2016}.
For FeSe$_{1-x}$S$_x$ near the nematic end point at $x \sim 0.17$
we find that a $H-T$ dependence collapses onto a single curve, as shown in Fig.~\ref{FigSM_MR_Scaling}(e).
Despite this, the energy scaling of magnetoresistance
used to described the antiferromagnetic critical point in Ref.~\cite{Hayes2016} is
not obeyed in the vicinity of the nematic end point in FeSe$_{1-x}$S$_x$, as detailed in Fig.~\ref{FigSM_MR_Scaling}(g-i).
This could be due to additional constrains to be included
 either to account for the nematoelastic coupling \cite{Paul2017}
and/or the effect of  impurities.
For example, a very dirty sample of FeSe$_{1-x}$S$_x$ close to
$x_{nom} \sim 0.18$ was recently suggested to obey $H-T$ scaling \cite{Licciardello2019MR}.

For reasons described above,
we propose a different approach to model the magnetoresistance data in the nematic state of FeSe$_{1-x}$S$_x$,
using a power law in magnetic fields given by $\rho_{xx}(H)= \rho_{0}(H) + b H^{\delta}$.
Strikingly, we find that all the magnetoresistance data inside the nematic phase can be described by a unique exponent
$\delta \sim 1.55(5)$ over a large field window, as shown by the colour  plot in Fig.\ref{fig3}(c) as well as in Figs.\ref{fig2}(c) and \ref{FigSM_Hscaling_1p55_all}(a-d).
A detailed method of the extraction of $\delta$ and its stability over a large temperature and field window
is shown in Fig.\ref{FigSM_FeSe_Hn_exponent}.
Furthermore, this gives $\delta \sim 2$ for samples in the tetragonal phase (see Fig.\ref{fig3}(c)).
Inside the nematic phase, the Fermi surface of FeSe$_{1-x}$S$_x$ distorts anisotropically \cite{Watson2015a,Coldea2017}
and an unusual type of scattering could become operational due to presence of hot and cold spots along certain directions \cite{Wang2019}.

\begin{figure*}[htbp]
	\centering
\includegraphics[trim={0cm 0cm 0cm 0cm}, width=0.8\linewidth,clip=true]{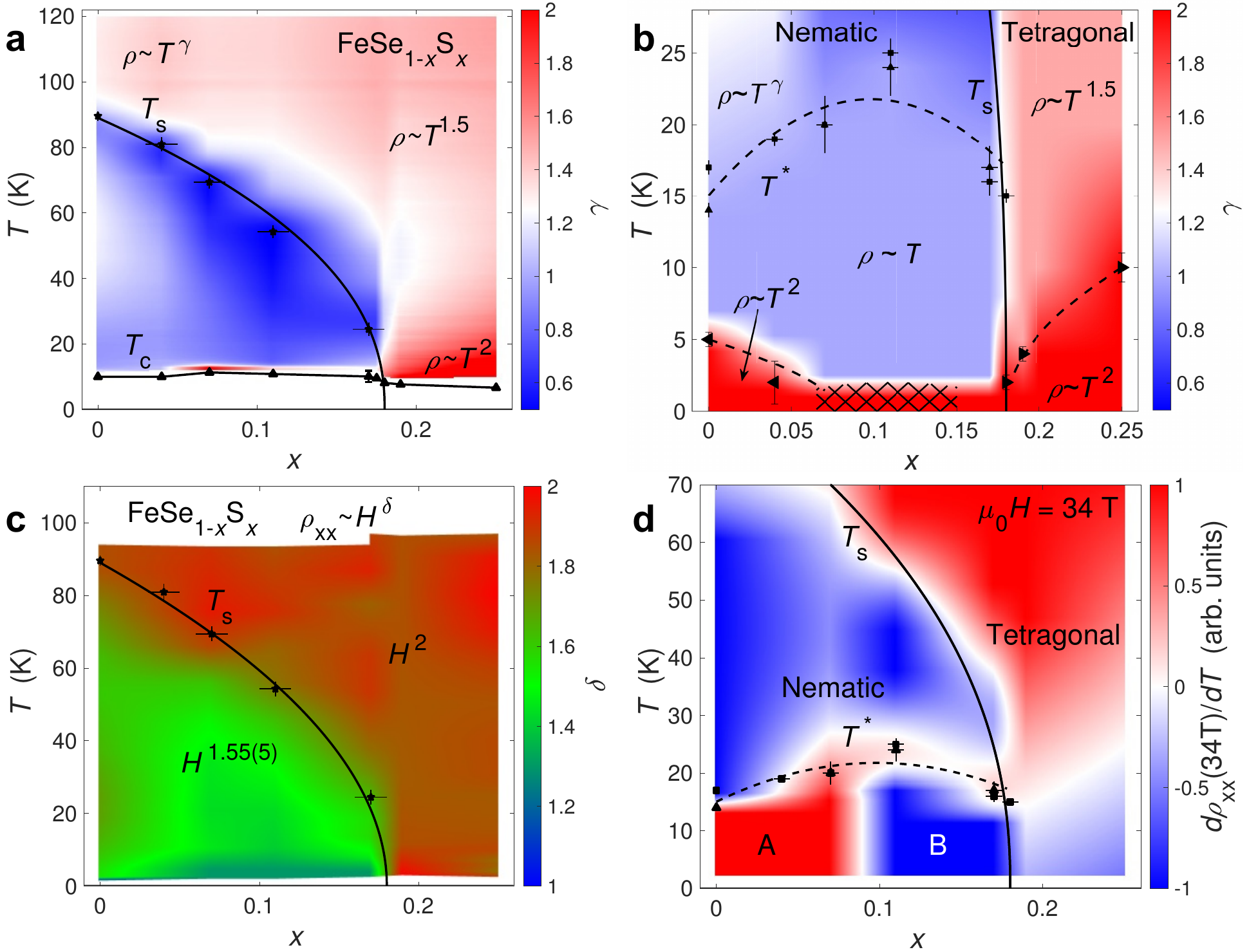}
	\caption{\textbf{Phase diagrams of the resistivity exponents and high-field transport in FeSe$_{1-x}$S$_x$.}
 The colour plot of the  temperature exponent,   $\gamma$, extracted from
(a)   zero-resistivity data, as shown in Fig.~\ref{FigSM_FeSe_Tn_exponent}.
(b) The low-temperature resistivity exponent below $T^*$, extrapolated from high magnetic fields
as shown in Figs.~\ref{FigSM_RvsB_Hparab_all} and \ref{FigSM_rho_vs_T_lowT_Hparac},
indicating the non-Fermi liquid behaviour of the nematic phase. Fermi liquid recovers
below $T_{FL}$ for the compositions $x$ with lowest disorder
both inside the nematic phase and in the tetragonal phase.
 (c) The  temperature dependence of field exponent $\delta$ showing a dominant $\sim H^{1.55}$ power law inside the nematic phase
 (based on Fig.\ref{FigSM_FeSe_Hn_exponent}).
 (d) The colour plot of the slope of resistivity in 34~T between the nematic A and B phases.
 Solid squares represent $T_{s}$ and solid triangles $T_c$. $T^*$ indicated by stars
represents the peak in magnetoresistance and the maximum in $|R_H|$.
 Solid lines indicate the nematic and superconducting phase boundaries and the dashed lines are guides to the eye.
 The hashed region at low temperatures in (b) has not yet been accessed experimentally.}
	\label{fig3}
\end{figure*}

In the absence of magnetic field the transport behaviour
can also be described by a power law, $\rho(T)=\rho_{0} +  A T^{\gamma}$.
Fig.~\ref{fig3}a shows a colour plot of the exponent $\gamma$,
which is close to unity at low temperatures inside the nematic phase and becomes sublinear close to the nematic phase boundary,
 indicating  a significant deviation from  Fermi-liquid behaviour
 (a value of $\gamma$=1.1(2) was previously reported for FeSe \cite{Kasahara2010}).
Outside the nematic phase  a $T^{1.5}$ dependence of resistivity
describes the data well over a large temperature range up to 120~K (see Fig.~\ref{fig2}(a) and Fig.\ref{fig3}(a)),
in agreement with previous studies of FeSe$_{1-x}$S$_x$ under pressure \cite{Reiss2019}.
Using the high-magnetic field data below $T_c$,
we extract the low-temperature resistivity in the absence of superconductivity, $\rho_{H \rightarrow 0}$(T).
Fig.~\ref{fig2}(d-f) shows resistivity against temperature for different values of $x$, together
with the extrapolated high-field points, using longitudinal magnetoresistance
when $H|| (ab)$ plane, shown in Fig.\ref{FigSM_RvsB_Hparab_all}.
We also use transverse magnetoresistance data to  extract the zero-field resistivity,
using  the established power law $H^{1.55}$, as shown in Fig.~\ref{FigSM_rho_vs_T_lowT_Hparac}.
From both measurements, we
find strong evidence for a linear resistivity in the low temperature regime, below $T^*$, inside the nematic  phase.
Linear resistivity was also detected  from the 35~T
 temperature dependence of the longitudinal magnetoresistance in Ref.\cite{Licciardello2019},
 however, it was assumed to occur near the {\it nematic critical point} defined as $x_{nom} \sim 0.16$,
 which corresponds to $x \sim 0.13$ in our phase diagrams in Fig.\ref{fig3} and Fig.\ref{FigSM_rho0_vs_x}(b)
 (as the resistivity derivative in Ref.\cite{Licciardello2019} show a $T_s \sim$ 51~K).
At low temperatures, we observe that Fermi-liquid behaviour recovers
in the tetragonal phase (see also Refs.~\cite{Urata2016,Licciardello2019})
 and inside the nematic phase, below $T_{FL}$ (see Figs.~\ref{fig2}(d-f) and \ref{fig3}(b)).
 This is  strongly dependent on composition and impurity level, even in the vicinity of
 the nematic end point (see Figs.~\ref{FigSM_rhoT_LowT_Hparab_FL_x18pc} and \ref{FigSM_extrapolation_comparison_dirty}).
 We find that $T_{FL}$  is highest for the samples
 with the largest residual resistivity ratio (above $\sim 16$)
(see Figs.\ref{FigSM_rho0_vs_x}(c) and \ref{FigSM_rhoT_LowT_Hparab_FL}).
Theoretical models suggest that the temperature exponent, $\gamma$,
in vicinity of critical points is highly dependent
on the presence of {\it cold spots} on different Fermi surfaces, due to the symmetry
of the nematic order parameter \cite{Wang2019,Maslov2011}.
On the other hand, near a antiferromagnetic critical point in the presence of spin fluctuations
the impurity level also affects the temperature exponent \cite{Rosch1999}.
Furthermore, the scale at which the crossover to Fermi liquid behavior occurs at $T_{FL}$
in nematic critical systems could depend on the strength of the coupling to the lattice \cite{Paul2017}.

An overall representation of the resistivity slope $d \rho_{xx}{\rm (34~T)}/dT$  in 34~T
for FeSe$_{1-x}$S$_x$ as a function of temperature is
shown in the phase diagram in Fig.~\ref{fig3}d.
The low-temperature manifestation of the nematic A and B phases is clearly
different below $T^*$. In order  to identify possible sources of scattering responsible for these changes,
we consider the role of spin fluctuations.
Recent NMR data found that anti-ferromagnetic spin fluctuations are present
inside the nematic phase of FeSe$_{1-x}$S$_x$,
being strongest around $x \sim 0.1$ \cite{Wiecki2018}.
In FeSe, spin fluctuations are rather anisotropic \cite{Cao2018,Wiecki2018}
and strongly field-dependent below 15~K  \cite{Shi2018}.
Interestingly, the spin-fluctuations relaxation rate is enhanced
 below $T^*$  (Fig.~\ref{fig3}(d)),
suggesting a correlation between spin-dependent
scattering, the high-field magnetoresistance and the low-temperature transport inside the nematic state.
High-magnetic fields are expected to align
magnetic spins and could affect the energy dispersion of low-energy spin excitations and  spin-dependent scattering in magnetic fields.
In FeSe, the spin-relaxation rate in different magnetic fields up to 19~T deviates at  $T^*$ \cite{Shi2018}
but it remains relatively constant in 19~T at the lowest temperatures.
This may suggest the variation in magnetoresistance in high magnetic fields at low temperatures in FeSe$_{1-x}$S$_x$
is more sensitive to the changes in the electronic behaviour rather to the spin fluctuations across
the nematic phase.

 The low-temperature regime below $T^*$ displays linear resistivity,
which is a potential manifestation of scattering induced by critical
spin-fluctuations in clean systems \cite{Rosch1999}.
 $\mu$SR studies place FeSe
near an itinerant antiferromagnetic quantum critical point at very low temperatures \cite{Grinenko2018}
and spin-fluctuations are only found inside the nematic state  \cite{Wiecki2018,Shi2018}.
On the other hand, close to the nematic end point in FeSe$_{1-x}$S$_x$
we find that resistivity is not linear in temperature
but is dominated by a $T^{1.5}$ dependence.
This is contrast to the linear resistivity found  near a antiferromagnetic critical point
in BaFe$_2$(As$_{1-x}$P$_x$) \cite{Kasahara2010}.
Theoretically, $\gamma=3/2$ could describe the resistivity
caused by strong antiferromagnetic critical fluctuations in the dirty limit \cite{Rosch1999,Moriya1985}.
However, in FeSe$_{1-x}$S$_x$ the spin fluctuations are suppressed and
a Lifshitz transition was detected at the nematic end point \cite{Coldea2019}.
At a nematic critical point the divergent fluctuations for different Fermi surfaces could display
unusual power laws in resistivity, as discussed in Refs.~\cite{Wang2019,DellAnna2007,Maslov2011}.
To asses the critical behaviour, it is worth emphasizing
 that the effective masses associated to the outer hole bands
do not show any divergence close to the nematic end point $x \sim 0.18$ \cite{Coldea2019}.
This agrees with the variation of the $A^{1/2}$ coefficient (see Fig.~\ref{FigSM_A_values})
and previous studies under pressure \cite{Reiss2019},
suggesting the critical nematic fluctuations
could be quenched by the coupling to the lattice along certain directions in FeSe$_{1-x}$S$_x$.

The striking difference in magnetotransport behaviour between the nematic and tetragonal phase in FeSe$_{1-x}$S$_x$
can have significant implications on what kind
of superconductivity is stabilized inside and outside the nematic phase
as different pairing channels may be dominant in different regions,
as found experimentally \cite{Sato2018,Hanaguri2018}.
Linear resistivity found  at low temperatures inside the nematic state is present
in the region where spin-fluctuations are likely to be present.
Furthermore, the absence of superconductivity enhancement  at the nematic end point  in FeSe$_{1-x}$S$_x$
is supported by the lack of divergent critical fluctuations,
found both with chemical pressure \cite{Coldea2019} and applied pressure \cite{Reiss2019}.
It is expected that the coupling to the relevant lattice strain 
restricts criticality in nematic systems only to certain high symmetry directions \cite{Labat2017,Paul2017}.

 In conclusion, we have studied the evolution of the
low-temperature magnetotransport behaviour in FeSe$_{1-x}$S$_x$ in high-magnetic fields up to 69~T.
We find that the nematic state has non-Fermi liquid behaviour
 and displays unconventional power laws in magnetic field,
reflecting the dominant anomalous scattering inside the nematic phase.
In high magnetic fields, well-above the upper critical fields,
the transverse magnetoresistance shows a change in slope that reflects the changes
in the spin-fluctuations and/or the electronic structure.
In the low-temperature limit, high magnetic field  suppresses superconductivity
and it reveals an extended linear resistivity in temperature
followed by a Fermi-liquid like dependence, highly dependent on the composition and impurity level.
Our study reveals the anomalous transport behaviour of the nematic state, strikingly different from the tetragonal phase,
 that influences how superconductivity is stabilized in different phases.

\vspace{-0.7cm}
\section{Materials and Methods}
\vspace{-0.5cm}
Single crystals of FeSe$_{1-x}$S$_x$
were grown by the KCl/AlCl$_3$ chemical vapor transport method \cite{Bohmer2013}.
The composition for samples from the same batch were checked using EDX
as reported previously in Ref.~\cite{Coldea2019}. Note that in Refs.\cite{Licciardello2019,Licciardello2019MR}
the nominal, $x_{nom}$ were can be at least 80\% less than the real $x$ (see also Ref.~\cite{Hosoi2016,Wiecki2018,Coldea2019}).
The structural transition at $T_s$  also provides useful information about the expected $x$ value,
as shown in Fig.\ref{FigSM_rho0_vs_x}.
More than 30 samples were screened for high magnetic field studies to test their
physical properties.
Residual resistivity ratio varies between 15-44, as shown in Fig.\ref{FigSM_rho0_vs_x}c.
We observed the variation within the same batch due to the inhomogeneous distribution of sulfur
with increasing $x$ (see Figs.\ref{FigSM_rho0_vs_x} and \ref{FigSM_rhoT_LowT_Hparab_FL_x18pc}).
We estimate that the nematic end point is located close to
$x \sim 0.180(5)$  (see Figs.\ref{FigSM_rho0_vs_x})  and  \ref{FigSM_A_values}).

In-plane transport measurements ($I||$({\it ab})) were performed in
a variable temperature cryostat in dc fields up to 38\,T at HFML, Nijmegen and up to 70~T at LNCMI, Toulouse
with the magnetic field applied mainly along the $c$-axis (transverse magnetoresistance)
but also in the ($ab$) conducting plane (longitudinal magnetoresistance)
at constant temperatures.
Low-field measurements were performed in a 16~T Quantum Design PPMS.
The resistivity $\rho_{xx}$ and Hall $\rho_{xy}$  components were
measured using a low-frequency  five-probe technique and were separated by (anti)symmetrizing data
measured in positive and negative magnetic fields. Good electrical contacts were achieved by In soldering along
the long edge of the single crystals and electrical currents up to 3~mA were used to avoid heating.
Magnetic fields along the $c$-axis  suppress
superconductivity in fields higher than 20~T for all $x$ values \cite{Coldea2019}.

\vspace{-0.7cm}
 \section{Acknowledgments}
 \vspace{-0.5cm}
 We thank Lara Befatto, Dmitrii  Maslov, Rafael Fernandes, Erez Berg, Shigeru Kasahara,
 Steve Simon, Siddharth Parameswaran and Stephen Blundell for useful comments and discussions.
  We thank and acknowledge previous contributions  from Matthew Watson, Mara Bruma,  Samuel Blake, Abhinav Naga and Nathaniel Davies.
 This work was mainly supported by  EPSRC  (EP/L001772/1,  EP/I004475/1,  EP/I017836/1).
  A.A.H. acknowledges the financial support of the Oxford Quantum
Materials Platform Grant (EP/M020517/1).
A portion of this work was performed at the National High Magnetic Field Laboratory, which is supported by National Science Foundation Cooperative Agreement No. DMR-1157490 and the State of Florida.
Part of this work was supported supported by
HFML-RU/FOM and LNCMI-CNRS, members of the European Magnetic Field
Laboratory (EMFL) and by EPSRC (UK) via its membership to the EMFL
(grant no. EP/N01085X/1).
Part of this work was supported by Programme Investissements d’ Avenir under the programme ANR-11-IDEX-0002-02, reference ANR-10-LABX-0037-NEXT
We also acknowledge the Oxford Centre for Applied Superconductivity
and the Oxford John Fell Fund.
A.I.C. acknowledges an EPSRC Career Acceleration Fellowship (EP/I004475/1).

\vspace{-0.7cm}
\section{Footnotes}
\vspace{-0.5cm}

To whom correspondence may be addressed: amalia.coldea@physics.ox.ac.uk

\vspace{0.3cm}
Author contributions:
A.I.C designed, planned and supervised the research.
M.B., P.R., Z.Z. and A.I.C. performed experiments in Nijmegen with support from A.M.;
P.R. and A.I.C. performed experiments in Tallahassee with support from D.G.;
M.B., P.R., and A.I.C. performed experiments in Toulouse with support from W.K.;
A.A.H., T.W. and S.S grew single crystals. M.B. and A.I.C. performed data analysis.
A.I.C and M.B. wrote the paper with contributions and comments from all the authors.

\bibliography{FeSeS_MR_bib_feb19}

\title{Supplemental Material}
\maketitle

\newcommand{\blue}{\textcolor{blue}}
\newcommand{\bdm}[1]{\mbox{\boldmath $#1$}}

\renewcommand{\thefigure}{S\arabic{figure}} 
\renewcommand{\thetable}{S\arabic{table}} 

\newlength{\figwidth}
\figwidth=0.48\textwidth

\setcounter{figure}{0}

\newcommand{\fig}[3]
{
\begin{figure}[!tb]
\vspace*{-0.1cm}
\[
\includegraphics[width=\figwidth]{#1}
\]
\vskip -0.2cm
\caption{\label{#2}
\small#3
}
\end{figure}}

\begin{table*}
\centering
\begin{tabular}{c}
{\bf \large Supplemental Materials for} \\
 {\bf \large Anomalous high-magnetic field  electronic state of the nematic superconductors Fe(Se$_{1-x}$S$_x$)}

\end{tabular}
\end{table*}

\begin{figure*}[ht]
	\centering
	\includegraphics[width=0.6\linewidth,clip=true]{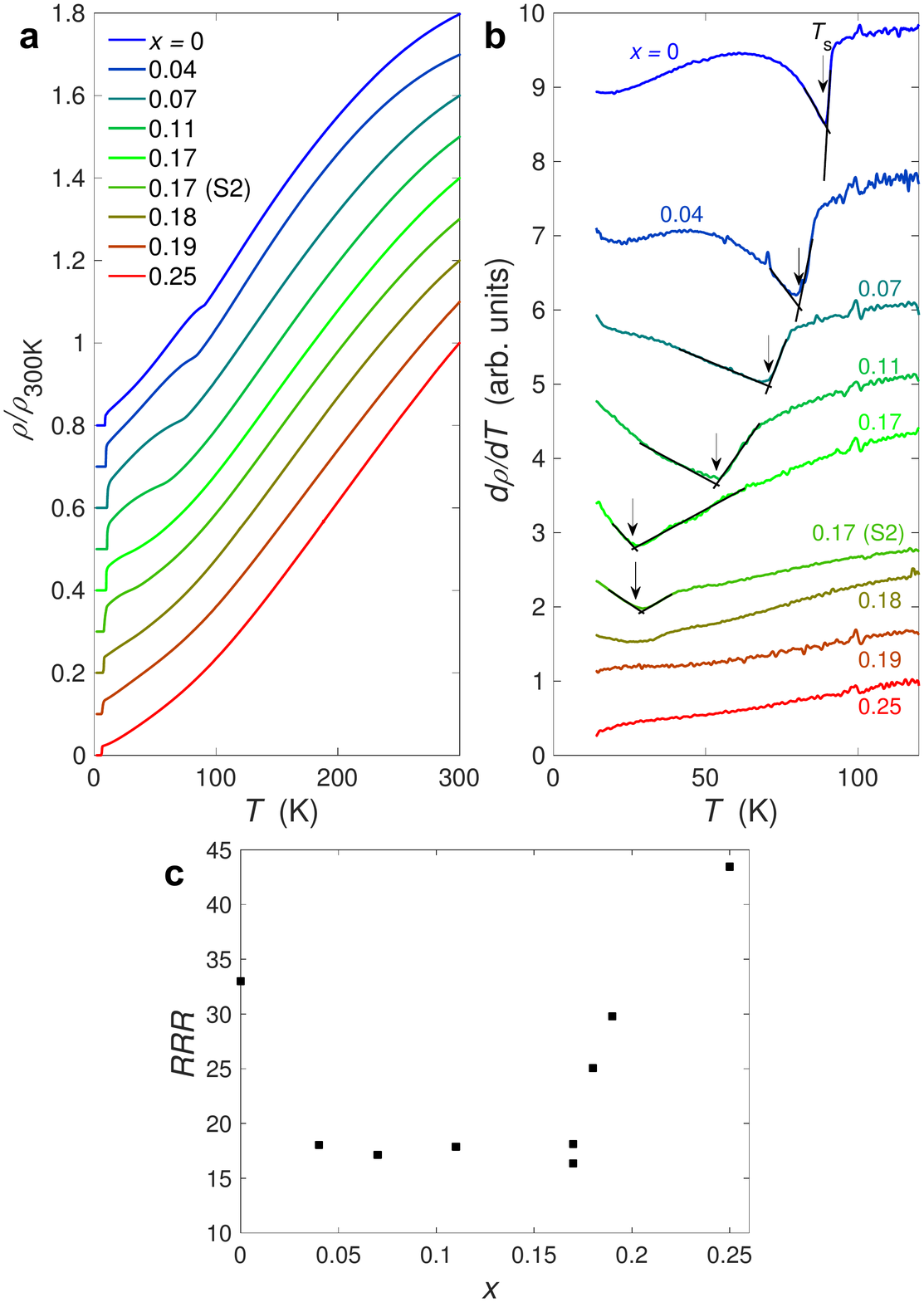}
	\caption{\textbf{Temperature dependence of the resistivity of FeSe$_{1-x}$S$_x$.}
		(a) Resistivity, normalised to the $300\,$K value, against temperature for different sulphur concentrations.
(b) The first derivative of the resistivity with respect to temperature for the same data.
The curves for different sulphur concentrations have been offset for clarity.
 The location of the structural transition, $T_{\rm{s}}$ is defined by the
 the intercept of the linear fits on either side of the transition, as indicated by arrows.
(c) The residual resistivity ratio (defined as the ratio between the room temperature and the onset of superconductivity resistivity), $RRR$, as a function of $x$.
The complete suppression of the structural transition occurs at $x_{\rm c} \sim 0.180(5) $, which agrees with previous reports \cite{Coldea2019,Hosoi2016}.
This value however differs from that reported in Ref. \cite{Licciardello2019}, where the nominal concentrations have been used.
For example, in Ref \cite{Licciardello2019}  $x_{\rm nom}=0.16$  has $T_{\rm{s}}\sim51$~K, which would correspond to $x \sim0.13$,
based on our phase diagram and previous reports \cite{Coldea2019,Hosoi2016}.
The two $x\sim0.17$ and the $x\sim0.18$ samples come from the same batch and their differences reflect the sulphur variation
and the degree of disorder ($x \sim 0.18$ is cleaner with an $RRR$  of $\sim24$ compared with $\sim16$ for the two $x\sim0.17$ samples).
 For $x\sim0.18$ the derivative in (b) evolves more gradually, without a well-defined structural
 transition as compared to the others, and we believe that this sample is the closest to the nematic end
 point (just inside the nematic state).}
	\label{FigSM_rho0_vs_x}
\end{figure*}

\begin{figure*}[htbp]
	\centering
	\includegraphics[width=0.99\linewidth,clip]{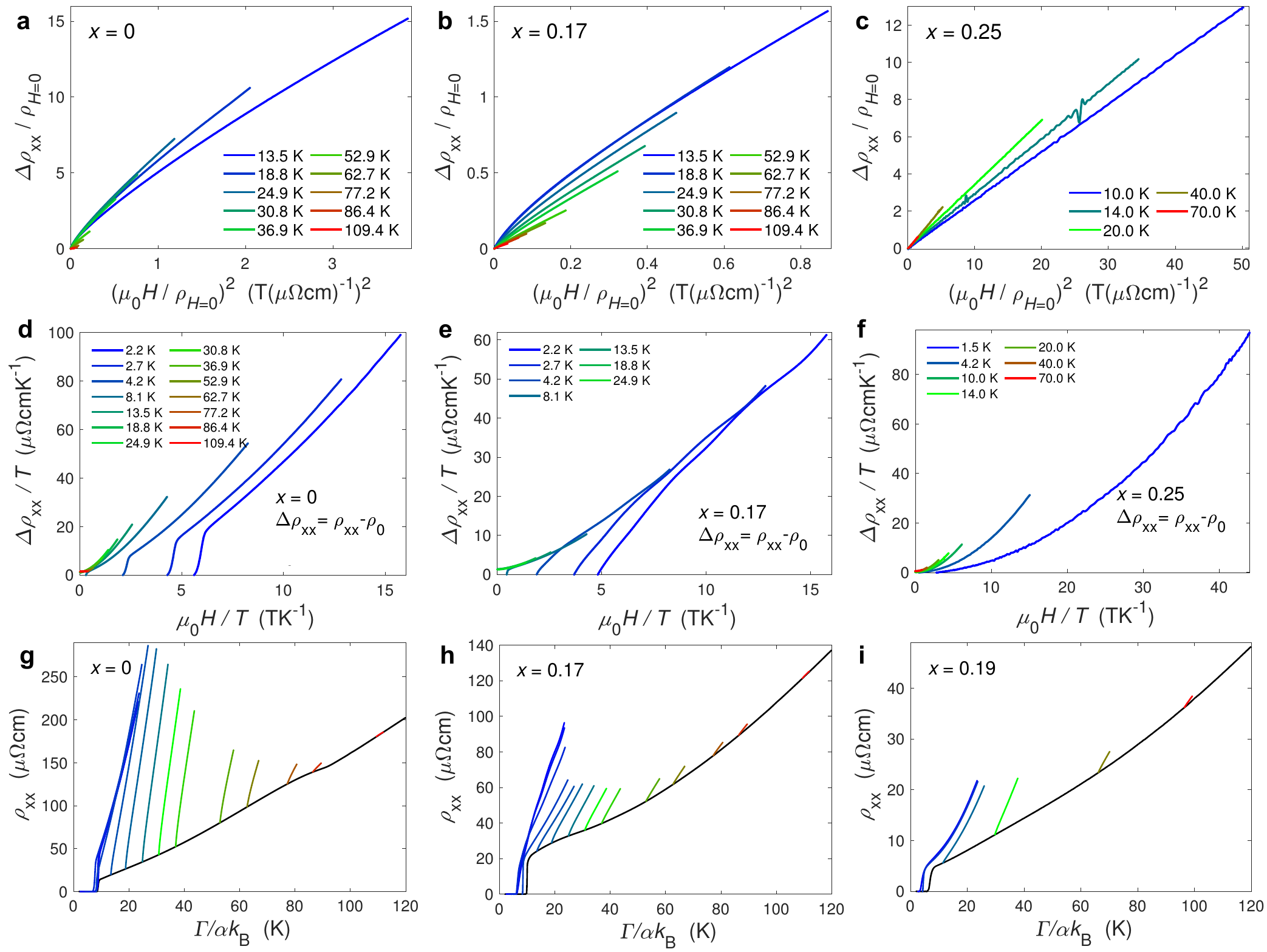}
	\caption{\textbf{The magnetoresistance scaling in  FeSe$_{1-x}$S$_{x}$.}
		(a-c) Kohler's plots, $\Delta \rho_{xx}(\mu_{\rm{0}}\it{H})/ \rho_{xx}(\it{H}\rm{=0}) \sim (\mu_{\rm{0}}\it{H}/\rho_{\rm{xx}}(\it{H}\rm{=0}))^{\rm{2}}$,
where $\Delta\rho_{\rm{xx}}=\rho_{\rm{xx}}-\rho_{\rm{xx}}(\it{H}\rm{=0})$ for $x=0,~ 0.17$ and $0.25$, respectively.
Kohler's rule is clearly violated for all samples over a large temperature range, providing evidence that electrical transport in these systems is not governed by a single scattering time.
		(d-f) $H -T$ scaling of $\Delta\rho_{\rm{xx}}/\rho_{\rm{0}} \sim \mu_{\rm{0}}\it{H}/\it{T}$, where $\Delta\rho_{\rm{xx}}=\rho_{\rm{xx}}-\rho_{\rm{0}}$ and $\rho_{\rm{0}}$ is the zero-temperature zero-field resistivity.
There is clearly no scaling for $x=0$ and $0.25$;
for the $x=0.17$ sample in the vicinity of the nematic end point,
data are dominated by a low-frequency quantum oscillation and collapse onto a single curve at low temperature, as shown in (e),
but we have not identified yet an appropriate scaling law for it.
However, in Ref.~\cite{Licciardello2019MR} for a particularly dirty sample of $x_{nom}=0.18$ with $RRR\sim5$ and
weak magnetoresistance  (a factor $\sim100$ smaller than our $x \sim 0.17$) $H/T$ scaling was proposed suggesting
that disorder may play an important role in this type of scaling in FeSe$_{1-x}$S$_{x}$.
(g-i) Energy scaling of resistivity as $\Gamma$, where $\Gamma=\alpha{}\it{k}_{\rm{B}}\it{T}\sqrt{1 + (\beta/\alpha)^{\rm{2}}(\mu_{\rm{B}}\mu_{\rm{0}}\it{H}/(\it{k}_{\rm{B}}\it{T}))^{\rm{2}}}$
using $\alpha=1$ and $\beta=1$, for $x=0, {}0.17$ and $0.19$, respectively.
 Our data do not follow the proposed energy scaling for any reasonable value of $\alpha/\beta$ and for any sulphur concentration in FeSe$_{1-x}$S$_{x}$.
This  magnetoresistance scaling was used to describe the antiferromagnetic critical region in BaFe$_2$(As$_{1-x}$P$_x$)$_2$ \cite{Hayes2016}. }
	\label{FigSM_MR_Scaling}
\end{figure*}

\begin{figure*}[htbp]
	\centering
	\includegraphics[trim={0cm 0cm 0cm 0cm}, width=0.99\linewidth,clip]{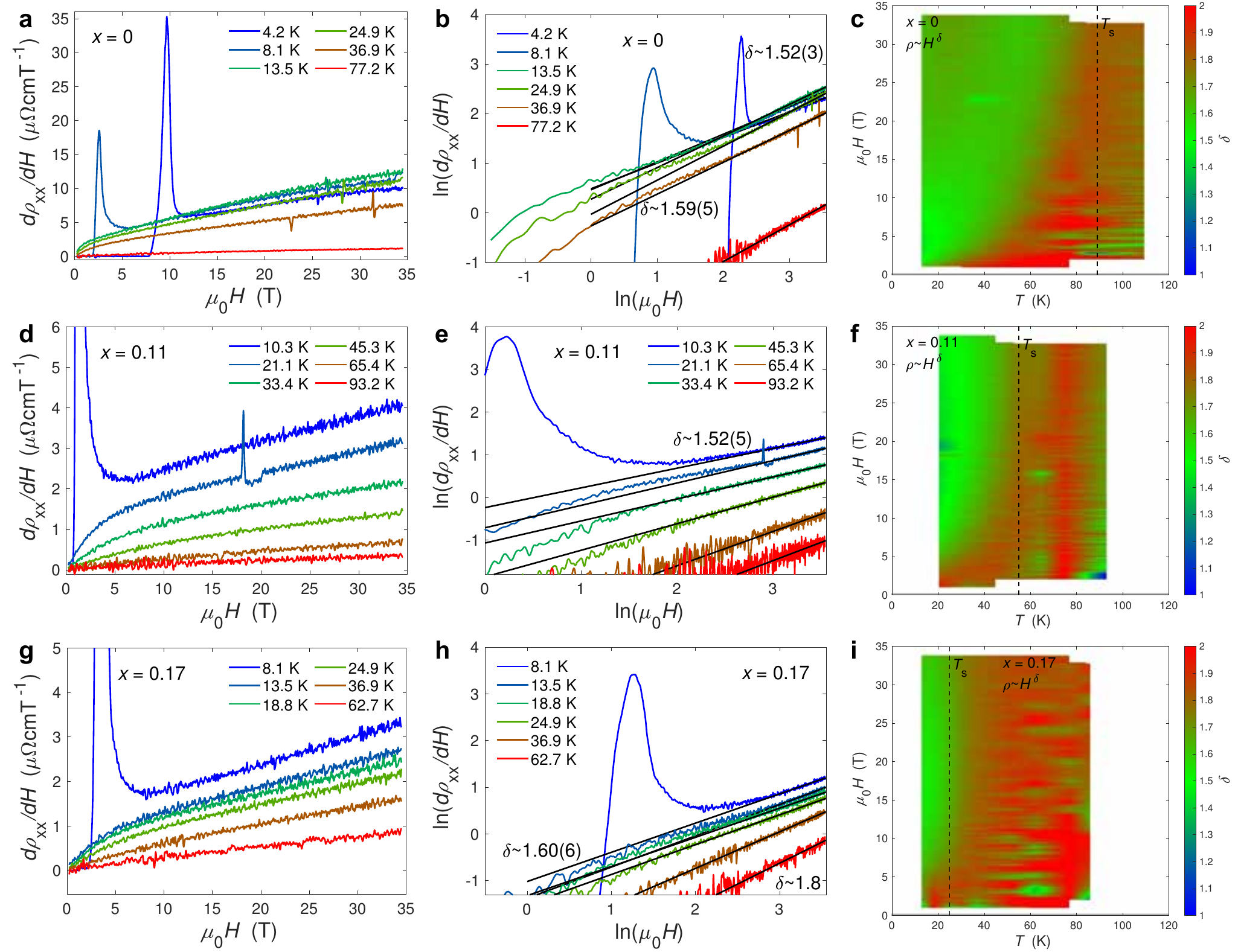}
	\caption{\textbf{Extraction of the magnetic field exponent $\delta$,  from the relationship $\rho_{\rm{xx}}=\rho_{H\rightarrow\rm{0}}+b H^{\delta}$, for FeSe$_{1-x}$S$_x$.}
		(a, d, g) Derivative of the symmetrized transverse resistivity with respect to the applied magnetic field for $x=0, {}0.11$ and $0.17$, respectively.
		(b, e, h) Natural logarithm of the derivative against the natural logarithm of magnetic field for the same data in (a, d, g).
Here linear fits to the high-field region allow the extraction of the magnetic field exponent from the gradient $\delta-1$.
		(c, f, i) Colour plots of $\delta$ in the temperature-magnetic field plane,
extracted above the superconducting transition temperature using the relationship $\it{d}\rm{ln}(\rho_{\rm{xx}}-	\rho_{xx}(\it{H}\rm{=0}))/\it{d}(\rm{ln}(\mu_{\rm{0}}\it{H})) = \delta$.
Consistent values of $\delta$ were obtained from linear fits over small regions using $\rm{ln}(\rho_{\rm{xx}}-\rho_{xx}(\it{H}\rm{=0})) = \rm{ln}(\it{b}) + \delta\rm{ln}(\mu_{\rm{0}}\it{H})$.
Our findings clearly show the evolution of the magnetic field exponent from  $\delta \sim 1.55(5)$ inside the nematic state towards $\delta \sim 2$ outside of the nematic state.}
	\label{FigSM_FeSe_Hn_exponent}
\end{figure*}

\begin{figure*}[htbp]
	\centering
		\includegraphics[trim={0cm 0cm 0cm 0cm}, width=1\linewidth,clip=true]{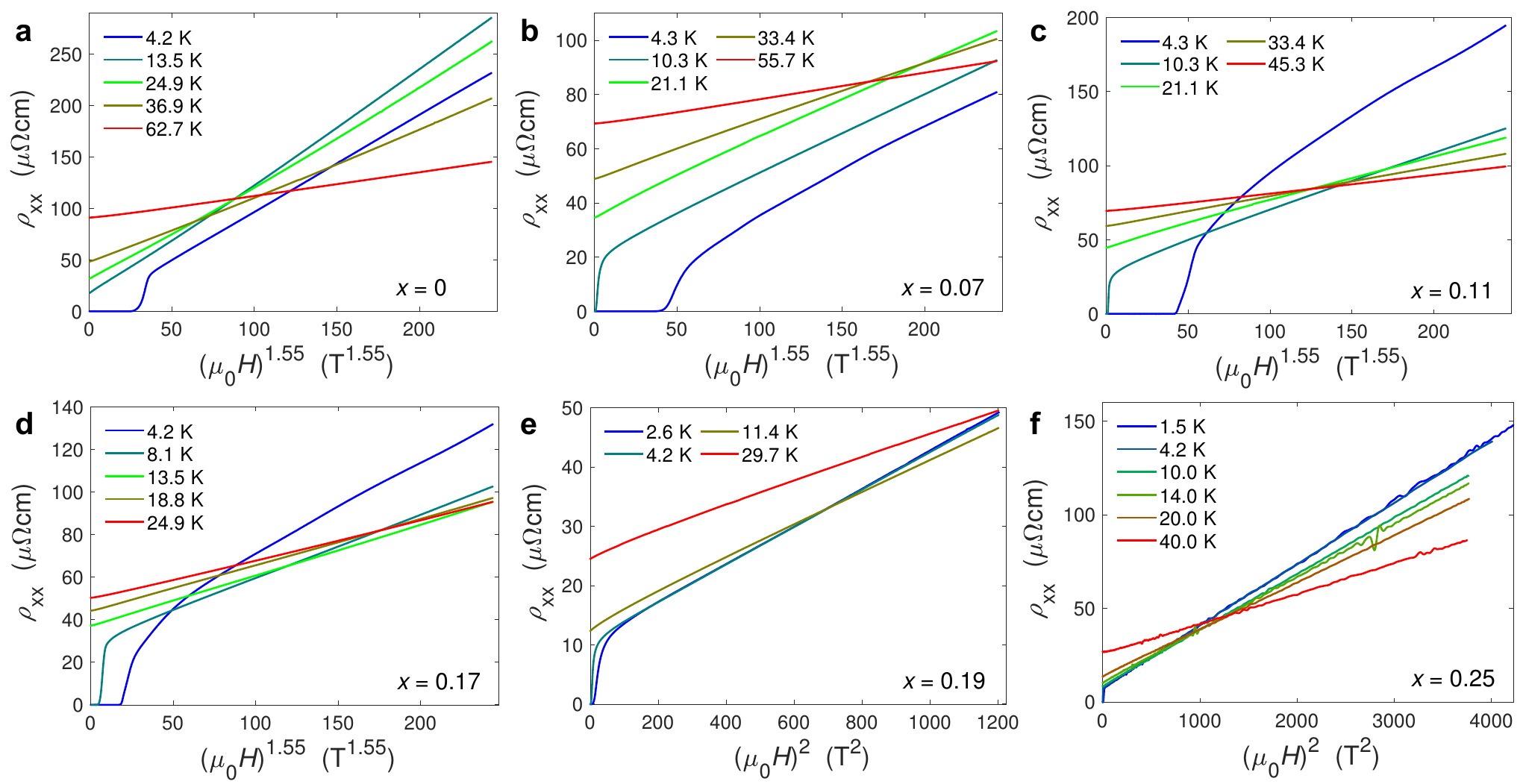}
	\caption{{\bf Field dependence of the transverse magnetoresistance of FeSe$_{1-x}$S$_x$ over a large magnetic field window.}
		(a-d) Resistivity as function of $(\mu_{\rm{0}}\it{H})^{\rm{1.55}}$ showing linear dependence
for samples inside the nematic phase at constant temperatures below $T_{\rm{s}}$
up to a magnetic field of $35\,$T.
At the lowest temperatures, below $\sim4.2\,$K, the magnetoresistance is dominated by quantum oscillations.
		(e-f) Resistivity versus $(\mu_{\rm{0}}\it{H})^{\rm{2}}$ at constant temperatures for samples in the tetragonal phase.
Data for $x=0.25$ were measured up to $\sim69\,$T.}
	\label{FigSM_Hscaling_1p55_all}
\end{figure*}

\begin{figure*}[htbp]
	\centering
	\includegraphics[width=1\linewidth,clip]{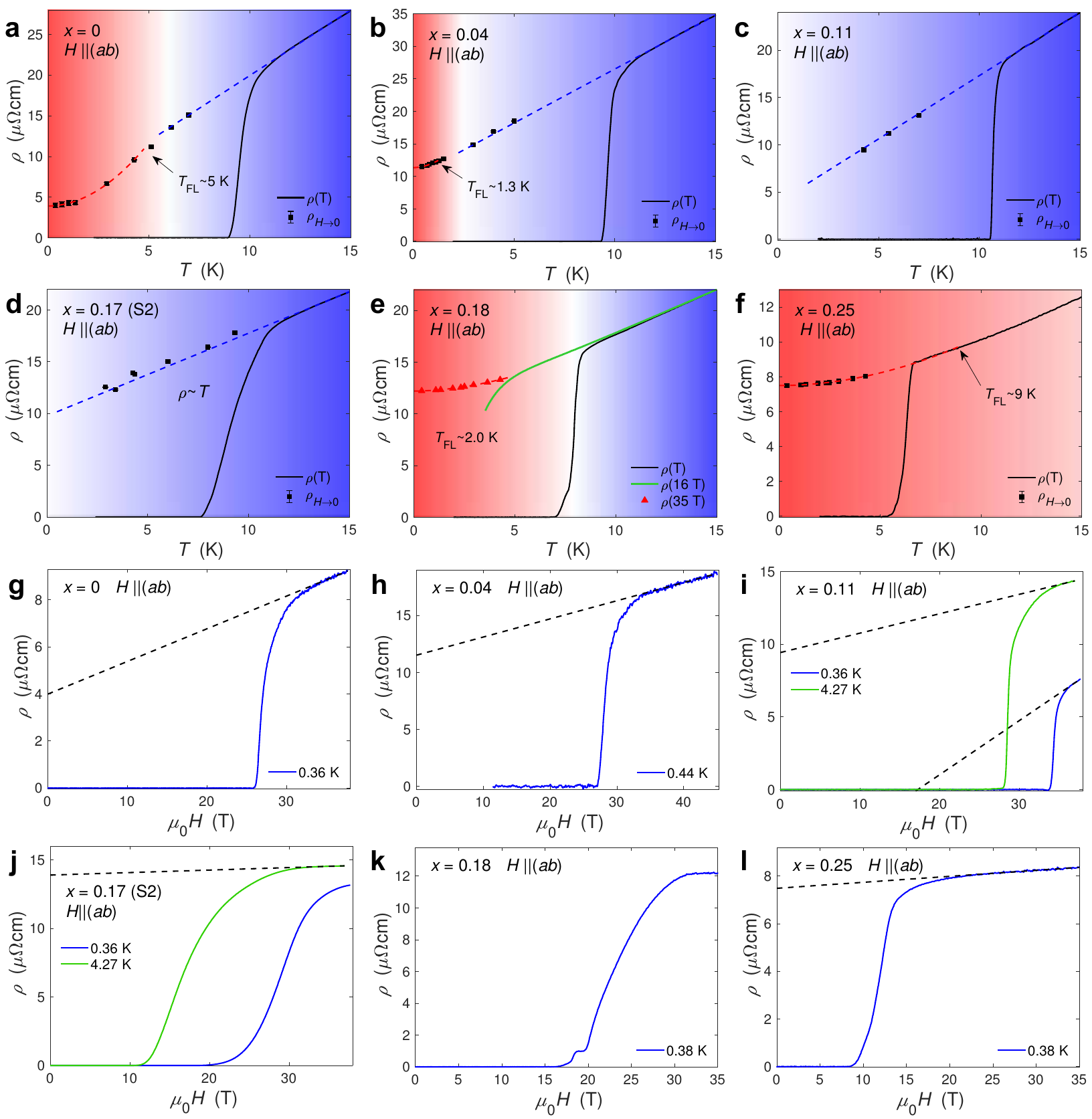}
	\caption{{\bf The low-temperature resistivity extracted from longitudinal magnetoresistance data of FeSe$_{1-x}$S$_x$.}
	(a-f) Temperature dependence of resistivity at low temperatures for different compositions used to build the low-temperature phase diagram in Fig.~3(b).
Solid squares show the extrapolated normal state resistivity, $\rho_{H \rightarrow 0 }$ from (g-l),
and the solid triangles in (e) are the resistivity data at 35~T from (k), when the magnetic field is along the conducting ($ab$) plane.
Solid lines are the zero-field resistivity for each sample.
Fermi-liquid like behaviour is observed in certain samples (with the largest resistivity ratio in Fig.\ref{FigSM_rho0_vs_x}(c))  below $T_{\rm{FL}}$, as indicated by arrows.
(g-l)	Magnetic field-dependence of the resistivity at the lowest temperature when $\it{H}||(\it{ab})$. Dashed lines are the linear extrapolation
towards $\it{H}\rightarrow{}0$.
 For $x=0.11$ and $0.17$, the upper critical field is too large to reach the normal state in magnetic fields up to $38\,$T in this orientation.}
	\label{FigSM_RvsB_Hparab_all}
\end{figure*}

\begin{figure*}[htbp]
	\centering
	\includegraphics[width=0.7\linewidth,clip=true]{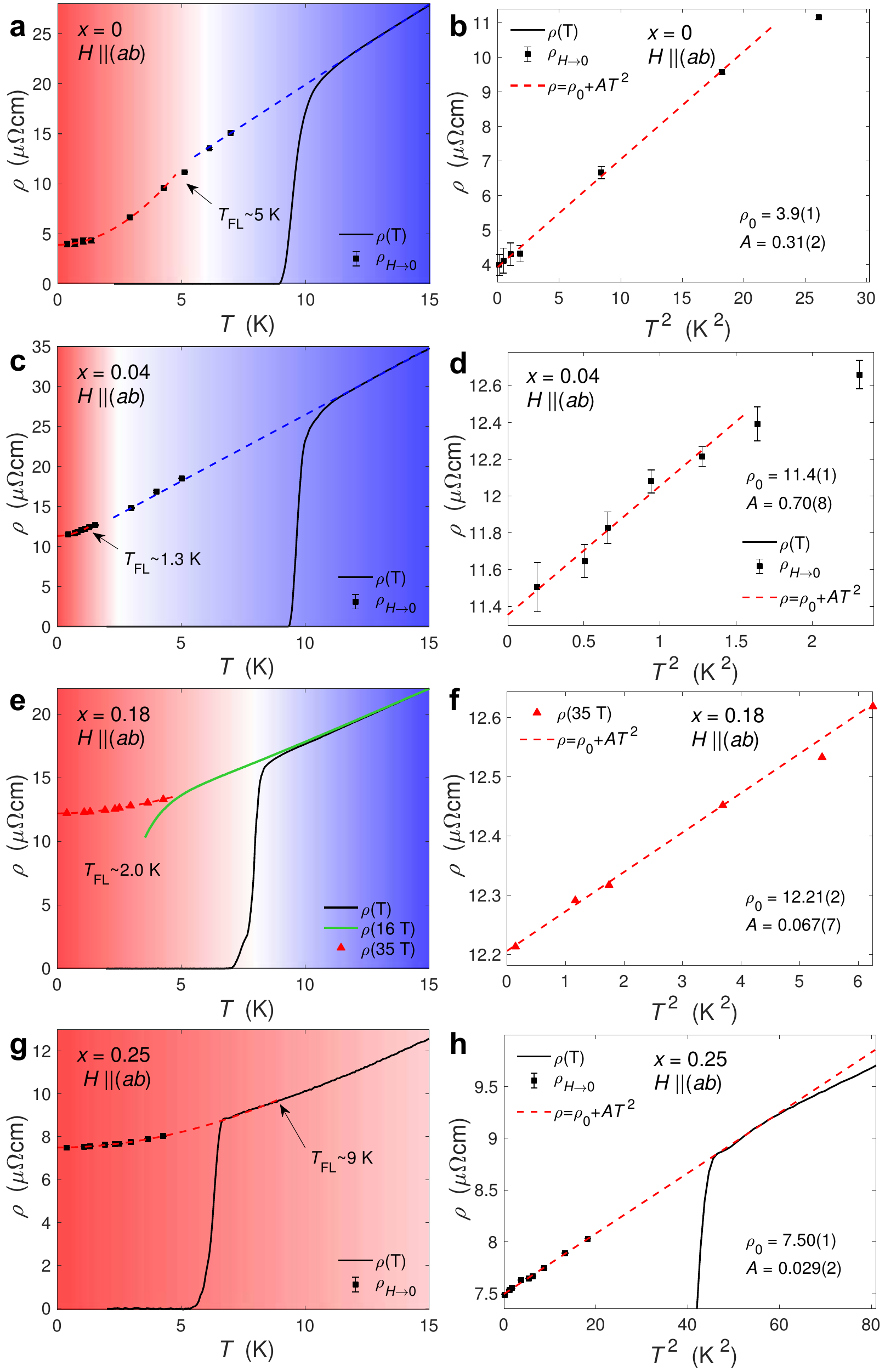}
	\caption{{\bf Evidence of Fermi liquid behaviour in certain samples of FeSe$_{1-x}$S$_x$.}
		(a, c, e, g) Temperature dependence of resistivity at low temperatures for different samples which show Fermi-liquid behaviour,
extracted as detailed previously in Fig.~\ref{FigSM_RvsB_Hparab_all}.
The red dashed lines are fits of resistivity to a quadratic temperature dependence below $T_{\rm{FL}}$, and the blue dashed lines
show a linear dependence between $\sim{}T_{\rm{FL}}$ and $T^{*}$. Resistivity data taken at $16\,$T for $x \sim 0.18 $ is also shown in (e)
and follows the zero field curve at high temperatures as expected for longitudinal magnetoresistance.
(b, d, f, h) The temperature dependence of resistivity against $T^{\rm{2}}$  illustrating the Fermi-liquid behaviour, given by $\rho=\rho_{\rm{0}}+AT^{\rm{2}}$.
Here the dashed red lines are linear fits in $T^{\rm{2}}$ and the zero-temperature resistivity values, $\rho_{\rm{0}}$, and $A$ parameters are listed in each panel.
We find that the samples with the larger $RRR$ also display larger  $T_{\rm{FL}}$.}
	\label{FigSM_rhoT_LowT_Hparab_FL}
\end{figure*}

\begin{figure*}[htbp]
	\centering
			\includegraphics[trim={0cm 0cm 0cm 0cm}, width=1\linewidth,clip]{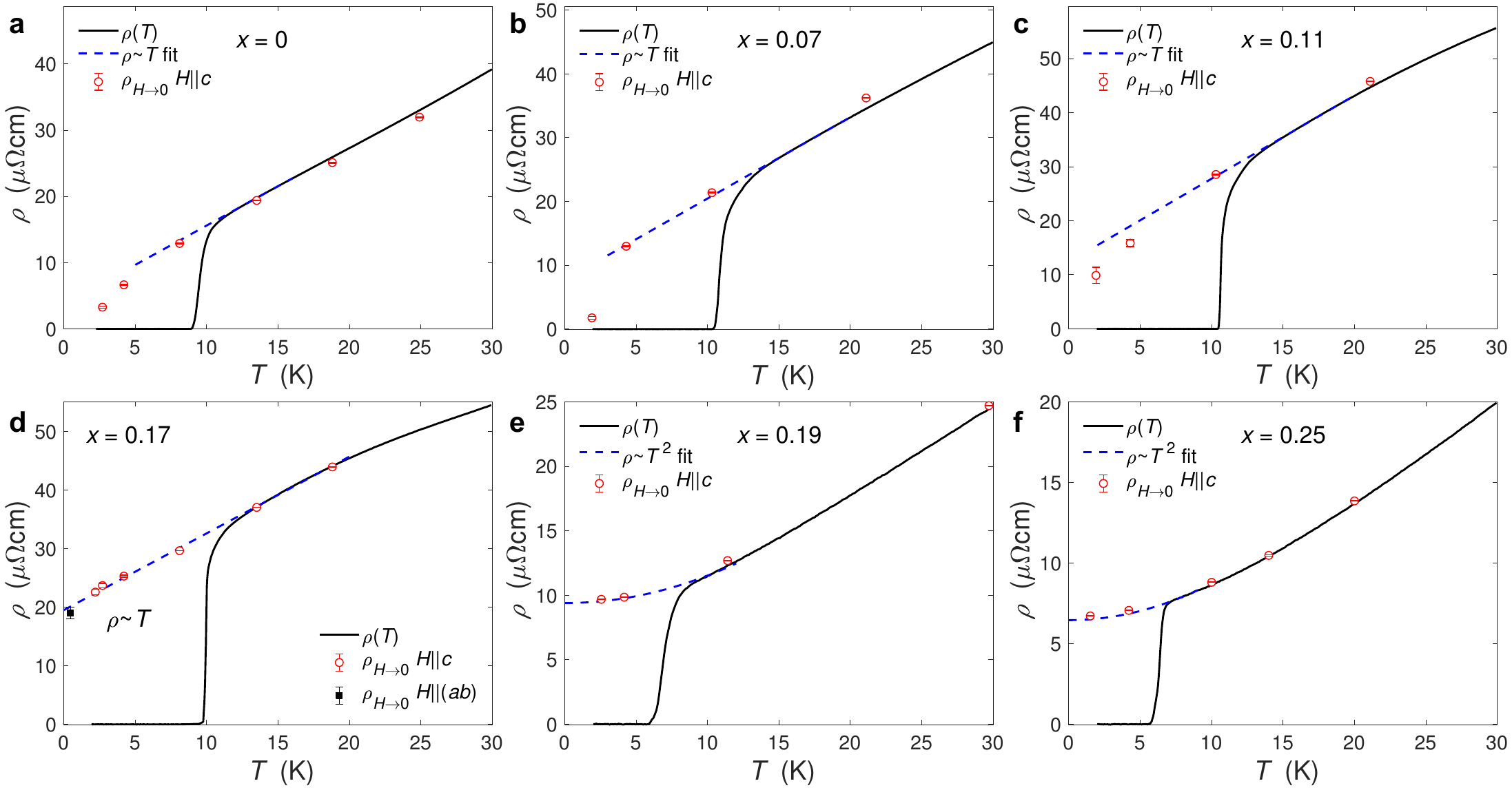}
	\caption{{\bf The low-temperature transport behaviour based on transverse magnetoresistance for FeSe$_{1-x}$S$_x$.}
		(a-f) Low temperature resistivity against temperature for different compositions.
The open circles in each panel correspond to the extrapolated resistivity (from symmetrized magnetic field data) up to $35\,$T,
using the magnetic field exponents of $\delta=1.55$ inside the nematic state ($x \lesssim  0.18$) and $\delta=2$
in the tetragonal state ($x=0.19-0.25$), as shown in Fig.~\ref{FigSM_Hscaling_1p55_all}.
Inside the nematic phase the presence of quantum oscillations at low temperatures,
makes the  extrapolation of $\rho_{\it{H}\rightarrow{}\rm{0}}$ more difficult in this orientation
when compared with longitudinal magnetoresistance studies, shown in Fig.~\ref{FigSM_RvsB_Hparab_all}.
In the tetragonal phase, Fermi-liquid behaviour is confirmed, similar to the longitudinal magnetoresistance studies in Fig.~\ref{FigSM_RvsB_Hparab_all}(f).
Dashed lines are either linear fits to the resistivity inside the nematic state or quadratic fits in the tetragonal state.
Linear resistivity is found in the vicinity of the nematic end point for $x \sim 0.17$ using both transverse and
longitudinal magnetoresistance studies.}
	\label{FigSM_rho_vs_T_lowT_Hparac}
\end{figure*}

\begin{figure*}[htbp]
	\centering
	\includegraphics[width=0.8\linewidth,clip=true]{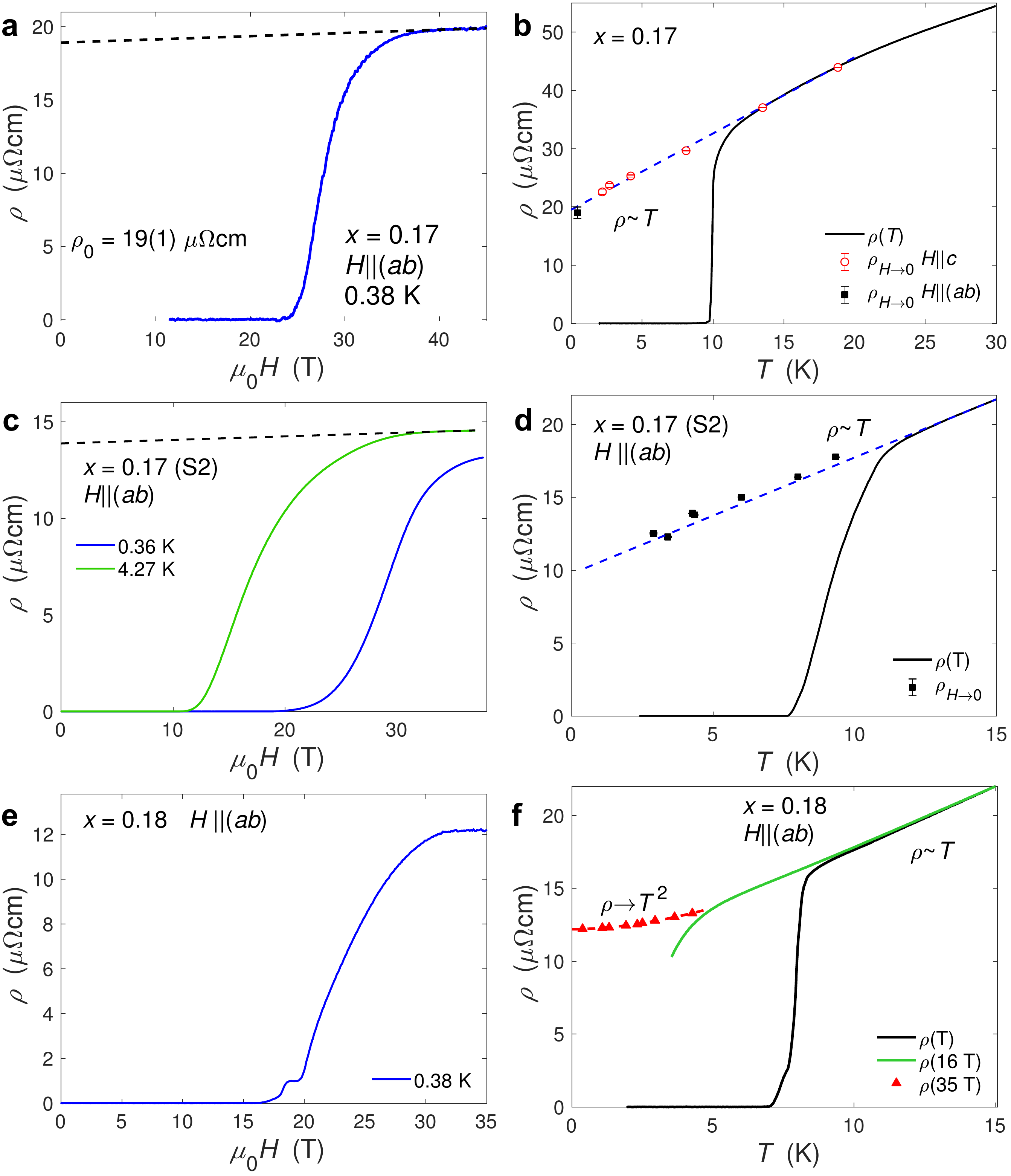}
	\caption{{\bf The low-temperature transport behaviour in the proximity
			of the nematic end point in FeSe$_{1-x}$S$_x$.}
The nematic end point  is defined as the
 complete suppression of the structural transition at $x_{\rm c} \sim 0.180(5) $, which agrees with previous reports \cite{Coldea2019,Hosoi2016}.
(a, c, e) Resistivity against magnetic field up to 45~T at the lowest temperatures ($\sim0.38\,$ K)  for three different samples from the same batch  with $\it{H}||(\it{ab})$.
For sample $x \sim 0.17 $ (S2)  extracting $\rho_{\it{H}\rightarrow{}\rm{0}}$ at the lowest temperatures is not possible up to 38~T and measurements at $\sim4.2\,$K, are also shown in (c).
			(b, d, f) Resistivity against temperature for the same samples.
Here the solid curves show the zero-field resistivity data and solid black squares
are the extrapolated $\rho_{\it{H}\rightarrow{}0}$ from $\it{H}||(\it{ab})$ field sweeps in (a, c).
Solid triangles in (f) are the resistivity at $35\,$T shown in (e).
Dashed blue lines in (b) and (d) show a linear dependence, and the red dashed line in (f) show a Fermi-liquid behaviour for $x \sim 0.18$.
The sample in which we find Fermi-liquid behaviour has a larger $RRR $ of $\sim24$ compared to $\sim16$ of the two other samples,
suggesting that Fermi-liquid behaviour  is observed only in cleaner samples.}
	\label{FigSM_rhoT_LowT_Hparab_FL_x18pc}
\end{figure*}

\begin{figure*}[htbp]
	\centering
	\includegraphics[trim={0cm 0cm 0cm 0cm}, width=1\linewidth,clip=true]{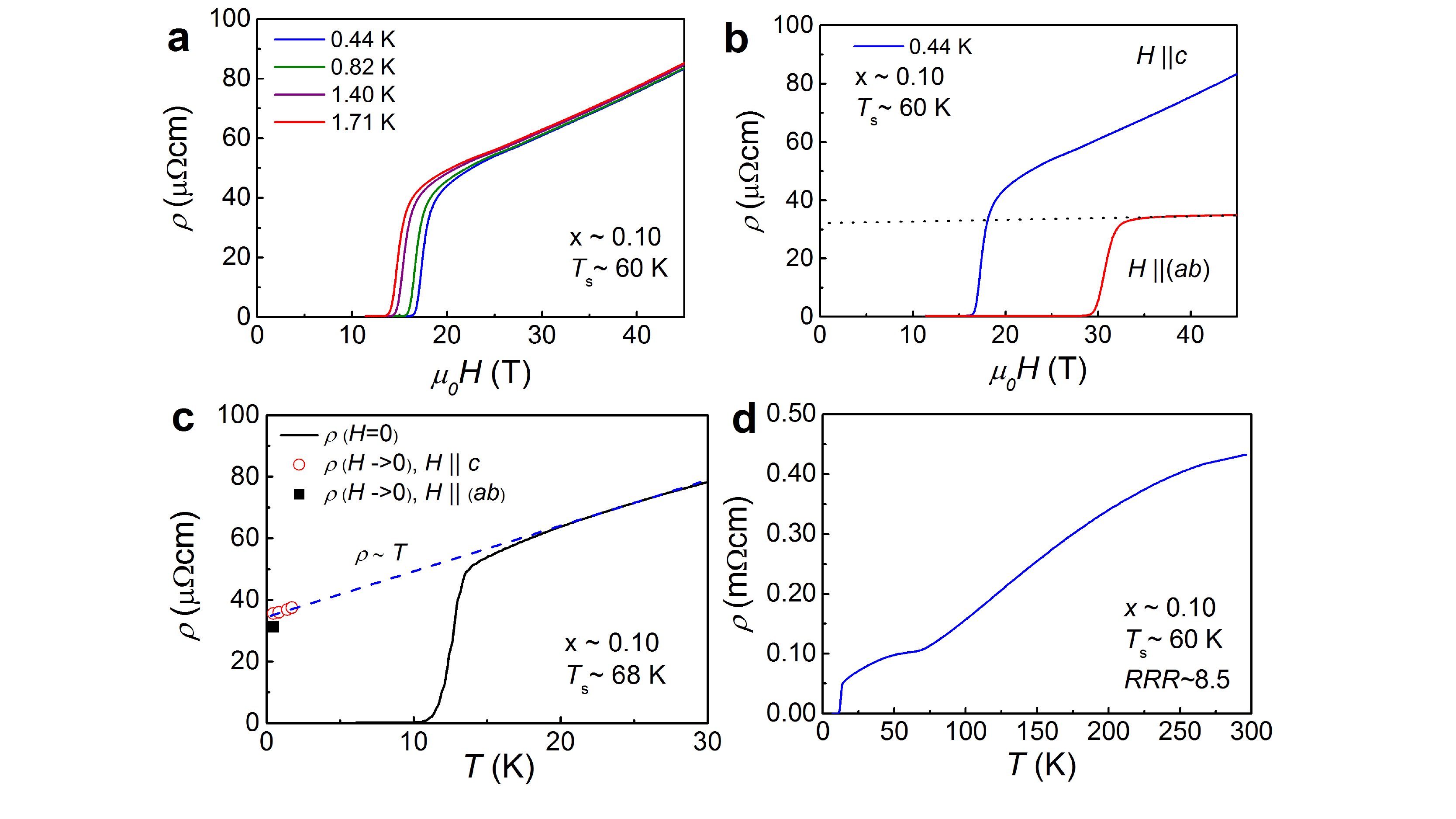}
	\caption{{\bf Magnetoransport data of a dirtier sample of FeSe$_{1-x}$S$_x$, with $x \sim 0.10$}.
		(a) Resistivity as a function of magnetic field at constant low temperatures with $H||c$.
		(b) Resistivity against magnetic field for two different orientations at base temperature, $H||c$ and $\it{H}||(\it{ab})$.
		(c) Zero-resistivity data (solid curve) together
with zero-field extrapolation using both linear extrapolation for $\it{H}||(\it{ab})$ (solid square) from (b)
and zero-field extrapolation using a $H^{1.55}$ dependence for $H||c$ (open circles) from (a).
		(d) Resistivity versus temperature for $x \sim 0.10$ with $RRR \sim 8.5$.
		The strong suppression of quantum oscillations and the low $RRR$ in this sample
are evidence that it is a dirtier system  which also display linear resistivity down to the lowest temperatures measured ($\sim 0.44$~K).}
	\label{FigSM_extrapolation_comparison_dirty}
\end{figure*}

\begin{figure*}[htbp]
	\centering
	\includegraphics[width=1\linewidth,clip]{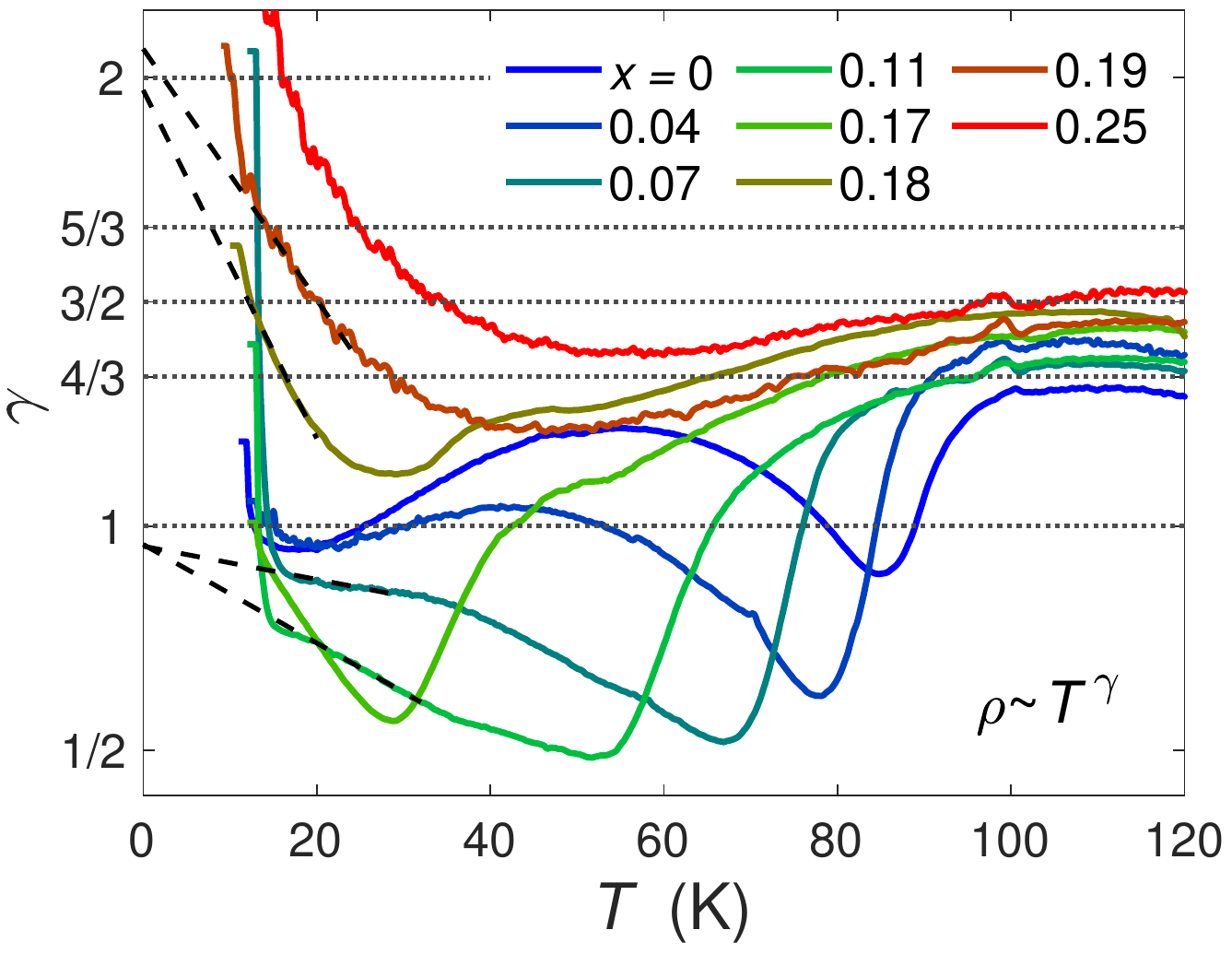}
	\caption{\textbf{The temperature dependence of the exponent $\gamma$ from $\rho=\rho_{\rm{0}}+\it{AT}^{\gamma}$ for FeSe$_{1-x}$S$_{x}$.}
		The $\gamma$ exponent is estimated a $\gamma=\it{d}\rm{ln}(\rho-\rho_{\rm{0}})/\it{d}\rm{ln}(\it{T})$ for various sulphur concentrations.
The $\rho_{\rm{0}}$ values are the zero-field zero-temperature resistivity extracted from Figs.~ \ref{FigSM_RvsB_Hparab_all}, \ref{FigSM_rhoT_LowT_Hparab_FL} and \ref{FigSM_rho_vs_T_lowT_Hparac}.
The curves shown here were used to generate the colour plots in Fig.~3(a) and (b) in the main body of the paper.
	Dashed black lines show possible extrapolations of the exponent towards the zero temperature limit.
The horizontal lines indicated different possible exponents predicted by theory in the vicinity of a critical point  \cite{Wang2019,DellAnna2007,Maslov2011}.
The $\gamma$ exponent  depends on the value of $\rho_{\rm{0}}$,
which was extracted at the lowest temperature from longitudinal magnetoresistance in Fig.~\ref{FigSM_RvsB_Hparab_all}.}
	\label{FigSM_FeSe_Tn_exponent}
\end{figure*}

\begin{figure*}[htbp]
	\centering
\includegraphics[trim={0cm 0cm 0cm 0cm}, width=1\linewidth,clip=true]{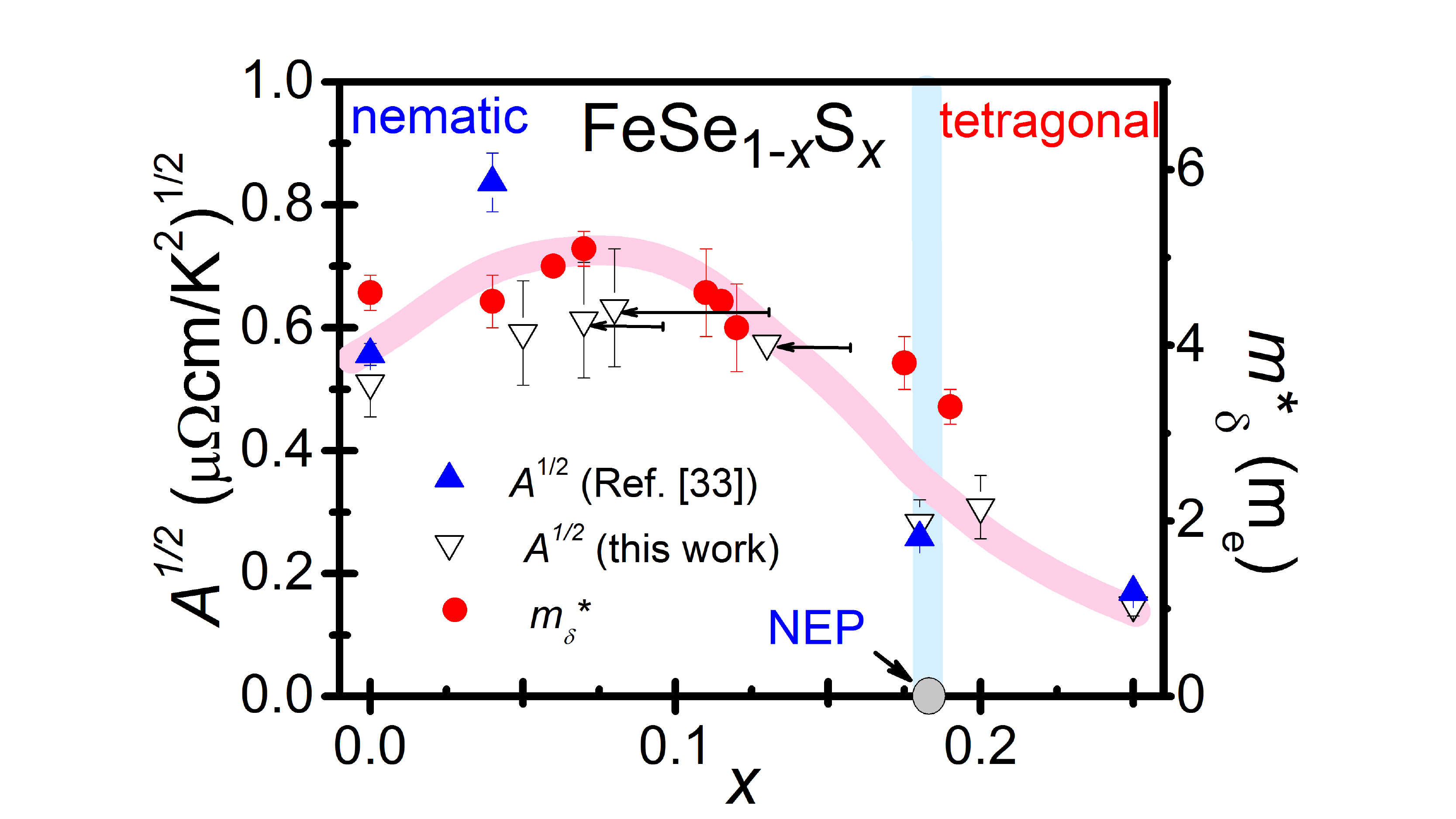}
	\caption{\textbf{Comparison between the $A^{1/2}$ temperature coefficient and the effective masses for different compositions of FeSe$_{1-x}$S$_x$.}
		The band masses of the $\delta$ orbit (from Ref.~\cite{Coldea2019})
are compared to the Fermi liquid coefficient, $A$, extracted at low temperatures (solid triangles) for compositions that show Fermi-liquid behaviour,
as shown in Fig.\ref{FigSM_rhoT_LowT_Hparab_FL}. Data reported in Ref.\cite{Licciardello2019} are also included for comparison as open triangles.
 Please note that Ref.\cite{Licciardello2019} uses nominal $x_{nom}$ values which are shifted to smaller values (as indicated by horizontal arrows) to match the
		real $x$ values based on EDX studies reported previously \cite{Coldea2019,Hosoi2016}. 
The nematic end point (NEP) is indicated by an arrow and circle around $x \sim 0.180(5)$.
Solid thick lines are guides to the eye.}
	\label{FigSM_A_values}
\end{figure*}

\end{document}